\documentstyle[prd,aps,preprint,tighten,epsfig]{revtex}

\begin{document}

\draft

\title{The T2K Indication of Relatively Large $\theta^{}_{13}$ and
a Natural Perturbation to the Democratic Neutrino Mixing Pattern}
\author{{\bf Zhi-zhong Xing}
\thanks{E-mail: xingzz@ihep.ac.cn}}
\address{Institute of High Energy Physics, Chinese Academy of Sciences,
Beijing 100049, China}

\maketitle

\begin{abstract}
The T2K Collaboration has recently reported a remarkable
indication of the $\nu^{}_\mu \to \nu^{}_e$ oscillation which is
consistent with a relatively large value of $\theta^{}_{13}$ in the
three-flavor neutrino mixing scheme. We show that
it is possible to account for such a result of $\theta^{}_{13}$
by introducing a natural perturbation to the democratic neutrino
mixing pattern, without or with CP violation.
A testable correlation between $\theta^{}_{13}$ and $\theta^{}_{23}$
is predicted in this ansatz. We also discuss the
Wolfenstein-like parametrization of neutrino mixing, and
comment on other possibilities of generating
sufficiently large $\theta^{}_{13}$ at the electroweak scale.
\end{abstract}


\newpage

Current solar, atmospheric, reactor and accelerator neutrino
oscillation experiments have provided us with very convincing evidence
that neutrinos are massive and lepton flavors are mixed \cite{PDG10}.
In the basis where the flavor eigenstates of three charged leptons
are identified with their mass eigenstates, the mixing of neutrino
flavors is effectively described by a $3\times 3$ unitary matrix $V$
whose nine elements can be parametrized in terms of three rotation
angles and three CP-violating phases:
\begin{eqnarray}
V = \left( \matrix{ c^{}_{12}
c^{}_{13} & s^{}_{12} c^{}_{13} & s^{}_{13} e^{-i\delta} \cr
-s^{}_{12} c^{}_{23} - c^{}_{12} s^{}_{13} s^{}_{23} e^{i\delta} &
c^{}_{12} c^{}_{23} - s^{}_{12} s^{}_{13} s^{}_{23} e^{i\delta} &
c^{}_{13} s^{}_{23} \cr s^{}_{12} s^{}_{23} - c^{}_{12} s^{}_{13}
c^{}_{23} e^{i\delta} & -c^{}_{12} s^{}_{23} - s^{}_{12} s^{}_{13}
c^{}_{23} e^{i\delta} & c^{}_{13} c^{}_{23} \cr} \right) P^{}_\nu \;
,
%     (1)
\end{eqnarray}
where $c^{}_{ij} \equiv \cos\theta^{}_{ij}$,
$s^{}_{ij} \equiv \sin\theta^{}_{ij}$ (for $ij = 12, 13, 23$),
and $P^{}_\nu ={\rm Diag}\{e^{i\rho}, e^{i\sigma}, 1\}$ is
a diagonal phase matrix which is physically relevant if
three neutrinos are the Majorana particles. The latest
global analysis of current neutrino oscillation data, done by
Schwetz {\it et al} \cite{Schwetz}, yields
$s^2_{12} = 0.312^{+0.017}_{-0.015}$,
$s^2_{13} = 0.010^{+0.009}_{-0.006}$ (NH) or $0.013^{+0.009}_{-0.007}$
(IH) and
$s^2_{23} = 0.51 \pm 0.06$ (NH) or $0.52 \pm 0.06$ (IH)
at the $1\sigma$ level, where ``NH" and ``IH"
correspond respectively
to the normal and inverted neutrino mass hierarchies. The central
values of three mixing angles are approximately
$\theta^{}_{12} \simeq 34^\circ$,
$\theta^{}_{13} \simeq 6^\circ$ and
$\theta{}_{23} \simeq 46^\circ$. Unfortunately,
three CP-violating phases of $V$ remain
entirely unconstrained. The ongoing and forthcoming neutrino
oscillation experiments will measure $\theta^{}_{13}$ and $\delta$,
and the neutrinoless double-beta decay experiments
will hopefully help to probe or constrain $\rho$ and $\sigma$.

\vspace{0.5cm}

The magnitude of $\theta^{}_{13}$ is one of the central concerns in
today's neutrino phenomenology. The most stringent upper bound on
this angle is $\theta^{}_{13} < 11.4^\circ$ at the $90\%$ confidence
level, as set by the CHOOZ \cite{CHOOZ} and MINOS \cite{MINOS}
experiments. Besides Ref. \cite{Schwetz}, there exist
several earlier analyses indicating that the
smallest neutrino mixing angle $\theta^{}_{13}$ might not be very small.
For example, $\theta^{}_{13} \simeq 7.3^{+2.0^\circ}_{-2.9^\circ}$
(1$\sigma$) by Fogli {\it et al} \cite{Fogli},
$\theta^{}_{13} \simeq 5.1^{+3.0^\circ}_{-3.3^\circ}$ ($1\sigma$)
by Gonzalez-Garcia {\it et al} \cite{GG},
and $\theta^{}_{13} \simeq 8.1^{+2.8^\circ}_{-4.5^\circ}$ as the best-fit
value by the KamLAND Collaboration \cite{KM}.
Although the statistical significance of these results remains quite low,
they {\it do} imply that $\theta^{}_{13}$ is possible to lie in the range
$5^\circ \lesssim \theta^{}_{13} \lesssim 11^\circ$.

\vspace{0.5cm}

A more robust indication of relatively large $\theta^{}_{13}$ comes
from the latest T2K measurement:
\begin{eqnarray}
0.03 < \sin^2 2\theta^{}_{13} < 0.28 & ~~~~ {\rm or} ~~~~ &
5.0^\circ \lesssim \theta^{}_{13} \lesssim 16.0^\circ ~~~~ ({\rm NH}) ,
\nonumber \\
0.04 < \sin^2 2\theta^{}_{13} < 0.34 & ~~~~ {\rm or} ~~~~ &
5.8^\circ \lesssim \theta^{}_{13} \lesssim 17.8^\circ ~~~~ ({\rm IH}) ,
%     (2)
\end{eqnarray}
for $\delta =0^\circ$ and at the $90\%$ confidence level \cite{T2K}. The
best-fit points are $\sin^2 2\theta^{}_{13} = 0.11$ (NH) or $0.14$ (IH),
corresponding to $\theta^{}_{13} = 9.7^\circ$ (NH) or $11.0^\circ$ (IH).
If such a value of $\theta^{}_{13}$ is finally established, it will
rule out a large number of neutrino mass models on the market
and provide us with a great hope to observe leptonic CP violation
in the long-baseline neutrino oscillation experiments in the foreseeable
future.

\vspace{0.5cm}

In this note we propose a phenomenologically
simple way to generate a sufficiently large value of $\theta^{}_{13}$. The
point is to introduce a natural perturbation to the
democratic neutrino mixing pattern $U$ \cite{FX96}, such that all three
mixing angles of $U$ receive comparable corrections which can be
as large as about $10^\circ$. We focus on a specific perturbation matrix
$X$ and determine its structure by using current experimental data on
the full neutrino mixing matrix $V = UX$. This ansatz predicts an interesting
correlation between $\theta^{}_{13}$ and $\theta^{}_{23}$, which
leads to $\theta^{}_{13} \simeq 9.6^\circ$ for $\theta^{}_{23}
= 45^\circ$, a result in good agreement with the
T2K indication. A Wolfenstein-like parametrization of $V$ and leptonic
CP violation are also discussed. Finally, we comment on a few
other possibilities of obtaining appreciable $\theta^{}_{13}$
at the electroweak scale.

\vspace{0.5cm}

Given a specific phase convention which will be convenient for our
subsequent discussions, the democratic mixing pattern reads
as follows \cite{FX96}:
\begin{equation}
U = \left( \matrix{ \displaystyle\sqrt{\frac{1}{2}} &
\displaystyle\sqrt{\frac{1}{2}}
& 0 \cr\vspace{-0.1cm}\cr
\displaystyle\sqrt{\frac{1}{6}} & -\displaystyle\sqrt{\frac{1}{6}} &
-\displaystyle\sqrt{\frac{2}{3}} \cr\vspace{-0.1cm}\cr
-\displaystyle\sqrt{\frac{1}{3}}
& \displaystyle\sqrt{\frac{1}{3}} &
-\displaystyle\sqrt{\frac{1}{3}} \cr} \right) \; ,
%     (3)
\end{equation}
whose three mixing angles are $\theta^{(0)}_{12} = 45^\circ$,
$\theta^{(0)}_{13} = 0^\circ$ and
$\theta^{(0)}_{23} =\arctan(\sqrt{2}) \simeq 54.7^\circ$ in
the standard parametrization as given in Eq. (1). It has been
pointed out that the tri-bimaximal
mixing pattern \cite{TB}, which is simply a ``twisted" form of the
democratic mixing pattern, can be directly obtained from $U$
by making an equal shift of its two nonzero mixing angles \cite{Xing2011}:
\begin{equation}
\theta^{}_* \equiv \theta^{(0)}_{12} - \vartheta^{(0)}_{12} =
\theta^{(0)}_{23} - \vartheta^{(0)}_{23} \simeq 9.7^\circ \; ,
%     (4)
\end{equation}
where $\vartheta^{(0)}_{12} = \arctan(1/\sqrt{2}) \simeq 35.3^\circ$
and $\vartheta^{(0)}_{23} = 45^\circ$ are the nonzero mixing angles of the
tri-bimaximal mixing pattern. Note that the value of $\theta^{}_*$
is quite suggestive because it is so close to the best-fit value of
$\theta^{}_{13}$ given by the present T2K data. Indeed, a novel and
viable neutrino mixing ansatz with
$\theta^{}_{13} \simeq \theta^{}_* \simeq 9.7^\circ$ has recently
been proposed in Ref. \cite{Xing2011}.

\vspace{0.5cm}

Note also that $U$ was originally obtained, as the leading term of
the lepton flavor mixing matrix $V$, from breaking the
$S(3)^{}_{\rm L} \times S(3)^{}_{\rm R}$ flavor symmetry of the
charged lepton mass matrix $M^{}_l$ in the basis where the neutrino
mass matrix $M^{}_\nu$ is diagonal \cite{FX96}. Here we assume
$V = U X$, where $X$ denotes a generic perturbation matrix which
can absorb small contributions from the flavor symmetry breaking
terms of both $M^{}_l$ and $M^{}_\nu$ \cite{FX96,Tanimoto,FX04}.
In general, of course, $U$ itself might come from either the charged
lepton sector or the neutrino sector, or both of them. The details
are certainly model-dependent.

\vspace{0.5cm}

To be explicit, we assume that $X$ has a simple pattern parallel to that
of $U$:
\begin{eqnarray}
X = \left( \matrix{
c^\prime_{12} & -s^\prime_{12} & 0 \cr
s^\prime_{12} c^\prime_{23} & c^\prime_{12} c^\prime_{23} & s^\prime_{23} \cr
s^\prime_{12} s^\prime_{23} & c^\prime_{12} s^\prime_{23} & -c^\prime_{23}
\cr} \right) \; ,
%     (5)
\end{eqnarray}
where $c^\prime_{ij} \equiv \cos\theta^\prime_{ij}$ and
$s^\prime_{ij} \equiv \sin\theta^\prime_{ij}$ (for $ij = 12, 23$).
The phase convention of $X$ is taken in such a way that all three
mixing angles of the full flavor mixing matrix $V= UX$ lie in the
first quadrant when CP is invariant. For simplicity, we tentatively ignore
possible CP-violating phases in $U$ and $X$. In this case we obtain
\begin{eqnarray}
V^{}_{e 1} & = &
\sqrt{\frac{1}{2}} \left( c^\prime_{12} + s^\prime_{12} c^\prime_{23}
\right) \; , \nonumber \\
V^{}_{e 2} & = &
\sqrt{\frac{1}{2}} \left( c^\prime_{12} c^\prime_{23} - s^\prime_{12}
\right) \; , \nonumber \\
V^{}_{e 3} & = &
\sqrt{\frac{1}{2}} \ s^\prime_{23} \; , \nonumber \\
V^{}_{\mu 3} & = &
\sqrt{\frac{1}{6}} \left( 2 c^\prime_{23} - s^\prime_{23} \right) \; ,
\nonumber \\
V^{}_{\tau 3} & = &
\sqrt{\frac{1}{3}} \left( c^\prime_{23} + s^\prime_{23} \right) \; ,
%     (6)
\end{eqnarray}
in which $\theta^\prime_{12}$ and $\theta^\prime_{23}$ are also assumed
to lie in the first quadrant. Comparing this result
with the standard parametrization of $V$ in Eq. (1), we immediately arrive at
\begin{eqnarray}
t^{}_{12} & = & \left|\frac{V^{}_{e2}}{V^{}_{e1}}\right|
= \frac{c^\prime_{12} c^\prime_{23} - s^\prime_{12}}
{c^\prime_{12} + s^\prime_{12} c^\prime_{23}} \; ,
\nonumber \\
s^{}_{13} & = & \left| V^{}_{e3} \right| =
\sqrt{\frac{1}{2}} \ s^\prime_{23} \; ,
\nonumber \\
t^{}_{23} & = & \left|\frac{V^{}_{\mu 3}}{V^{}_{\tau 3}} \right|
= \frac{2 c^\prime_{23} - s^\prime_{23}}
{\sqrt{2} \left( c^\prime_{23} + s^\prime_{23} \right)} \; ,
%     (7)
\end{eqnarray}
where $t^{(\prime)}_{ij} \equiv \tan\theta^{(\prime)}_{ij}$ (for $ij = 12, 23$).
Therefore,
\begin{eqnarray}
t^\prime_{23} & = & \frac{\sqrt{2} \left( \sqrt{2} - t^{}_{23}\right)}
{1 + \sqrt{2} \ t^{}_{23}} \; ,
\nonumber \\
t^\prime_{12} & = & \frac{1 + \sqrt{2} \ t^{}_{23} - t^{}_{12}
\sqrt{5 - 2\sqrt{2} \ t^{}_{23} + 4 t^2_{23}}}
{t^{}_{12} \left( 1 + \sqrt{2} \ t^{}_{23} \right)
+ \sqrt{5 - 2\sqrt{2} \ t^{}_{23} + 4 t^2_{23}}} \; .
%     (8)
\end{eqnarray}
Since both $\theta^{}_{13}$ and $\theta^{}_{23}$ depend on a
single parameter $\theta^\prime_{23}$, they have the following
correlation:
\begin{eqnarray}
s^{}_{13} = \frac{\sqrt{2} - t^{}_{23}}
{\sqrt{5 - 2\sqrt{2} \ t^{}_{23} + 4 t^2_{23}}} \; .
%     (9)
\end{eqnarray}
This expression can be regarded as the analytical prediction of our ansatz.
Some discussions about the above results are in order.
\begin{itemize}
\item     Given $\theta^{}_{23} = 45^\circ$, Eq. (9) leads us to
a numerical prediction of the smallest neutrino mixing angle
$\theta^{}_{13}$:
\begin{eqnarray}
\theta^{}_{13} = \arcsin\left[\frac{\sqrt{2} - 1}
{\sqrt{9 - 2\sqrt{2}}}\right] \simeq 9.6^\circ \; .
%     (10)
\end{eqnarray}
This result is in good agreement with the best-fit value of
$\theta^{}_{13}$ extracted from the T2K data. If $\theta^{}_{23}
\simeq 46^\circ$ is taken \cite{Schwetz}, one then arrives at
$\theta^{}_{13} \simeq 8.6^\circ$.

\item     Fixing $\theta^{}_{23} = 45^\circ$, we obtain
$\theta^\prime_{23} \simeq 13.6^\circ$ from Eq. (8). This
value is very close to the Cabibbo angle $\theta^{}_{\rm C}
\simeq 13^\circ$ of quark flavor mixing
\cite{PDG10}, whose sine function $\sin\theta^{}_{\rm C} \simeq
0.22$ can be treated as a perturbation to the
identity matrix to get the realistic Cabibbo-Kobayashi-Maskawa
matrix \cite{FX99}. Taking $\theta^{}_{12} \simeq 34^\circ$ together with
$\theta^{}_{23} = 45^\circ$, we can also obtain
$\theta^\prime_{12} \simeq 10.2^\circ$. It is interesting to see that
$\theta^\prime_{12}$ and $\theta^\prime_{23}$ are comparable in
magnitude, and they are also comparable with $\theta^{}_{13}$.
In this sense, we argue that the perturbation to $U$ is quite natural.

\item     If one simply assumes $\theta^\prime_{12} \simeq \theta^\prime_{23}
\simeq \theta^{}_{\rm C}$ from a model-building point of view at
the electroweak scale, then Eq. (7) gives the predictions
\begin{eqnarray}
\theta^{}_{12} & = & \arctan\left[\frac{\cos^2\theta^{}_{\rm C}
- \sin\theta^{}_{\rm C}}{\cos\theta^{}_{\rm C} \left( 1 +
\sin\theta^{}_{\rm C} \right)}\right] \simeq 31.3^\circ \; ,
\nonumber \\
\theta^{}_{13} & = & \arcsin\left[\sqrt{\frac{1}{2}} \
\sin\theta^{}_{\rm C}\right] \simeq 9.2^\circ \; ,
\nonumber \\
\theta^{}_{23} & = & \arctan\left[\frac{2\cos\theta^{}_{\rm C} -
\sin\theta^{}_{\rm C}}{\sqrt{2} \left( \cos\theta^{}_{\rm C} +
\sin\theta^{}_{\rm C} \right)}\right] \simeq 45.5^\circ \; ,
%     (11)
\end{eqnarray}
which are also consistent with current experimental data. An
explicit neutrino mass model of this nature will be explored elsewhere.

\item     The above hypothesis is interesting in the sense that it
suggests a Wolfenstein-like parametrization of the
neutrino mixing matrix \cite{Xing03}.
Setting $s^\prime_{12} = s^\prime_{23} =
\sin\theta^{}_{\rm C} \equiv \lambda \simeq 0.22$, we approximately
obtain
\begin{eqnarray}
V = \left( \matrix{
\displaystyle\sqrt{\frac{1}{2}} \left( 1 + \lambda \right) & \displaystyle\sqrt{\frac{1}{2}} \left( 1 - \lambda \right) &
\displaystyle\sqrt{\frac{1}{2}} \ \lambda \cr\vspace{-0.1cm}\cr
\displaystyle\sqrt{\frac{1}{6}} \left( 1 - \lambda \right) & -\displaystyle\sqrt{\frac{1}{6}} \left( 1 + 3 \lambda \right) &
\displaystyle\sqrt{\frac{2}{3}} \left( 1 - \frac{1}{2} \lambda \right) \cr\vspace{-0.1cm}\cr
-\displaystyle\sqrt{\frac{1}{3}} \left( 1 - \lambda \right) &
\displaystyle\sqrt{\frac{1}{3}} & \displaystyle\sqrt{\frac{1}{3}}
\left( 1 + \lambda \right) \cr} \right)
+ {\cal O}(\lambda^2) + \cdots .
%     (12)
\end{eqnarray}
It becomes transparent that eight of the nine matrix elements of
$U$ receive the ${\cal O}(\lambda)$ corrections. In other words,
all three mixing angles of $U$ get corrected in a quite similar way
and with a quite similar strength.
\end{itemize}
As pointed out in Ref. \cite{Xing2011}, it is
difficult to generate relatively large $\theta^{}_{13}$ from
{\it natural} perturbations to the tri-bimaximal mixing pattern,
unless the perturbations are adjusted in such a
way that its two nonzero mixing angles are slightly modified
but its vanishing mixing angle is significantly modified.
This kind of perturbations seem to be strange.

\vspace{0.5cm}

Now let us look at the possibility of introducing leptonic CP
violation into the neutrino mixing matrix $V$. For this purpose,
one of the simplest ways is to make the transformation
$s^\prime_{12} \to s^\prime_{12} e^{i\phi}$ with $\phi$ being a
real phase parameter. In this case $X$ becomes complex and
thus $V = UX$ contains a nontrivial CP-violating phase. Then
\begin{eqnarray}
V^{}_{e 1} & = &
\sqrt{\frac{1}{2}} \left( c^\prime_{12} + s^\prime_{12} c^\prime_{23}
e^{i\phi} \right) \; , \nonumber \\
V^{}_{e 2} & = &
\sqrt{\frac{1}{2}} \left( c^\prime_{12} c^\prime_{23} - s^\prime_{12}
e^{i\phi} \right) \; ;
%     (13)
\end{eqnarray}
and $V^{}_{\mu 1}$, $V^{}_{\mu 2}$, $V^{}_{\tau 1}$ and $V^{}_{\tau 2}$
are also complex. We obtain
\begin{eqnarray}
t^{}_{12} & = & \left|\frac{V^{}_{e2}}{V^{}_{e1}}\right| =
\frac{\sqrt{\left(c^\prime_{23}\right)^2 + \left(t^\prime_{12}\right)^2
- 2 t^\prime_{12} c^\prime_{23} \cos\phi}}
{\sqrt{1 + \left(t^\prime_{12} c^\prime_{23}\right)^2
+ 2 t^\prime_{12} c^\prime_{23} \cos\phi}} \; ,
\nonumber \\
J^{}_V & \equiv & {\rm Im}\left(V^{}_{e2} V^{}_{\mu 3} V^*_{e3} V^*_{\mu 2}\right)
= \frac{1}{6} c^\prime_{12} \left(s^\prime_{12}\right)^2
\left( 2 c^\prime_{23} - s^\prime_{12} \right) \left( c^\prime_{23}
+ s^\prime_{23} \right) \sin\phi \; ,
%     (14)
\end{eqnarray}
where $J^{}_V$ is the Jarlskog invariant of leptonic CP
violation. Note that the results for $s^{}_{13}$ and $t^{}_{23}$
are the same as those in Eq. (7), and thus Eq. (9) also holds in the present
ansatz. Typically taking $\theta^{}_{12} \simeq 34^\circ$ and $\theta^{}_{23}
= 45^\circ$, we first obtain $\theta^\prime_{23} \simeq 13.6^\circ$
from Eq. (8) and then the constraint equation
\begin{eqnarray}
\left(t^\prime_{12}\right)^2 - 4.96 t^\prime_{12} \cos\phi
+ 0.86 \simeq 0 \;
%     (15)
\end{eqnarray}
from Eq. (14). In the assumption of $\cos\phi \simeq 0.9$, for instance, we
arrive at $\theta^\prime_{12} \simeq 11.4^\circ$. The leptonic
Jarlskog invariant turns out to be $J^{}_V \simeq
4.8 \times 10^{-3}$, about two orders of magnitude larger than
the corresponding Jarlskog parameter in the quark sector \cite{FX99}.
Larger CP-violating effects are possible in this ansatz if one
assumes $\phi$ to be reasonably large, but
$\phi \simeq 90^\circ$ is forbidden as one can easily
see from Eq. (15). Because $J^{}_V = c^{}_{12} s^{}_{12}
c^2_{13} s^{}_{13} c^{}_{23} s^{}_{23} \sin\delta$ holds in the
standard parametrization of $V$, it is straightforward to establish
the relationship between $\delta$ and $\phi$ with the help of Eq. (14).

\vspace{0.5cm}

In the presence of CP violation as introduced above, the
Wolfenstein-like parametrization of $V$ in Eq. (12) becomes
\begin{eqnarray}
V = \left( \matrix{
\displaystyle\sqrt{\frac{1}{2}} \left( 1 + \lambda e^{i\phi} \right) & \displaystyle\sqrt{\frac{1}{2}} \left( 1 - \lambda e^{i\phi} \right) &
\displaystyle\sqrt{\frac{1}{2}} \ \lambda \cr\vspace{-0.1cm}\cr
\displaystyle\sqrt{\frac{1}{6}} \left( 1 - \lambda e^{i\phi} \right) & -\displaystyle\sqrt{\frac{1}{6}} \left( 1 + 2 \lambda
+ \lambda e^{i\phi} \right) &
\displaystyle\sqrt{\frac{2}{3}} \left( 1 - \frac{1}{2} \lambda \right) \cr\vspace{-0.1cm}\cr
-\displaystyle\sqrt{\frac{1}{3}} \left( 1 - \lambda e^{i\phi} \right) &
\displaystyle\sqrt{\frac{1}{3}} \left( 1 - \lambda + \lambda e^{i\phi} \right)
& \displaystyle\sqrt{\frac{1}{3}} \left( 1 + \lambda \right) \cr} \right)
+ {\cal O}(\lambda^2) + \cdots .
%     (16)
\end{eqnarray}
An appreciable value of $\theta^{}_{13}$ is also a good news to the
leptonic unitarity triangles \cite{FX99}, which can be used to geometrically
describe CP violation in the lepton sector. The area of each
unitarity triangle is equal to $|J^{}_V|/2 \simeq \lambda^2 |\sin\phi|/6$.
If the T2K experiment is finally able to probe the CP-violating asymmetry
between the probabilities of $\nu^{}_\mu \to \nu^{}_e$ and
$\overline{\nu}^{}_\mu \to \overline{\nu}^{}_e$ oscillations, then it will
be possible to determine $J^{}_V$ itself through
\begin{eqnarray}
P(\overline{\nu}^{}_\mu \to \overline{\nu}^{}_e) -
P(\nu^{}_\mu \to \nu^{}_e) = 16 J^{}_V
\sin \frac{\Delta m^2_{21} L}{4E} \sin \frac{\Delta m^2_{31} L}{4E}
\sin \frac{\Delta m^2_{32} L}{4E} \;
%     (17)
\end{eqnarray}
in the neglect of terrestrial matter effects. Even the matter
effects are non-negligible or significant,
it is likely to reconstruct the leptonic
unitarity triangles in vacuum from those effective ones in matter
and then pin down the genuine effect of CP violation \cite{Zhang}.

\vspace{0.5cm}

In summary, we have taken account of the robust T2K indication of
a relatively large value of $\theta^{}_{13}$ and paid particular
attention to how to confront a constant neutrino mixing
pattern, which may be motivated by a certain flavor symmetry and can
predict $\theta^{}_{13} =0^\circ$ in the symmetry limit, with
$\theta^{}_{13} \sim 10^\circ$. We have shown that a natural
perturbation to the democratic mixing pattern
$U$ can easily produce the realistic neutrino
mixing matrix $V$ with sufficiently large $\theta^{}_{13}$.
An interesting relationship between $\theta^{}_{13}$ and $\theta^{}_{23}$
has been predicted in this ansatz, and a Wolfenstein-like 
parametrization of $V$ has been discussed. We have also shown 
that it is possible for such an ansatz to accommodate leptonic 
CP violation, and its phenomenological consequences will soon be 
tested in a variety of more accurate neutrino oscillation experiments.

\vspace{0.5cm}

Generating $\theta^{}_{13} \sim 10^\circ$ from $\theta^{}_{13} = 0^\circ$
is certainly a very nontrivial job. Besides an explicit perturbation to
a given constant flavor mixing pattern like $U$, one may also consider
finite quantum corrections to $\theta^{}_{13}$ at the electroweak scale
\cite{Araki} or renormalization-group running effects on $\theta^{}_{13}$
from a superhigh-energy scale down to the electroweak scale \cite{RGE}.
However, it is in general difficult (if not impossible) for both approaches
to generate a sufficiently large value of $\theta^{}_{13}$, and
in particular $\theta^{}_{12}$ is usually most sensitive to radiative
corrections.

\vspace{0.5cm}

Of course, one may not necessarily start from $\theta^{}_{13} \sim 0^\circ$
for model building. For example,
the so-called tetra-maximal neutrino mixing pattern \cite{Xing08}
yields $\theta^{}_{12} = \arctan(2-\sqrt{2}) \simeq 30.4^\circ$,
$\theta^{}_{13} = \arcsin[(\sqrt{2}-1)/(2\sqrt{2})] \simeq 8.4^\circ$,
$\theta^{}_{23} = 45^\circ$ and $\delta = 90^\circ$ in the symmetry
limit. Hence this pattern can easily fit current experimental
data if one introduces slight corrections to it. The open question is
how to incorporate such a constant mixing scenario with a natural neutrino
mass model, and a possible answer to this question will be explored elsewhere.

\vspace{0.5cm}

{\it Acknowledgments:}
I would like to thank H.S. Chen for calling my particular attention
to the latest T2K result, and T. Araki, H. Fritzsch, Y.F. Li and J. Tang
for some interesting discussions. I am also grateful to A. Richter for 
his warm hospitality at the ECT* in Trento, where this paper was written.
This work was supported in part by the National Natural
Science Foundation of China under grant No. 10875131.


\newpage

\begin{thebibliography}{99}

\bibitem{PDG10} Particle Data Group, K. Nakamura {\it et al.},
J. Phys. G {\bf 37}, 075021 (2010).

\bibitem{Schwetz} T. Schwetz, M. Tortola, and J.W.F. Valle,
arXiv:1103.0734.

\bibitem{CHOOZ} The CHOOZ Collaboration, M. Apollonio {\it et al.},
Eur. Phys. J. C {\bf 27}, 331 (2003).

\bibitem{MINOS} The MINOS Collaboration, P. Adamson {\it et al.},
Phys. Rev. D {\bf 82}, 051102 (2010).

\bibitem{Fogli} G.L. Fogli, E. Lisi, A. Marrone, A. Palazzo,
and A.M. Rotunno, Phys. Rev. Lett. {\bf 101}, 141801 (2008).

\bibitem{GG} M.C. Gonzalez-Garcia, M. Maltoni, and J. Salvado,
JHEP {\bf 1004}, 056 (2010).

\bibitem{KM} The KamLAND Collaboration, A. Gando {\it et al.},
Phys. Rev. D {\bf 83}, 052002 (2011).

\bibitem{T2K} The T2K Collaboration, K. Abe {\it et al.},
arXiv:1106.2822.

\bibitem{FX96} H. Fritzsch and Z.Z. Xing,
Phys. Lett. B {\bf 372}, 265 (1996);
Phys. Lett. B {\bf 440}, 313 (1998); Phys. Rev. D {\bf 61}, 073016 (2000).

\bibitem{TB} P.F. Harrison, D.H. Perkins, and W.G. Scott, Phys.
Lett. B {\bf 530}, 167 (2002); Z.Z. Xing, Phys. Lett. B {\bf 533}, 85 (2002);
P.F. Harrison and W.G. Scott, Phys. Lett. B {\bf 535} (2002);
X.G. He and A. Zee, Phys. Lett. B {\bf 560}, 87 (2003).

\bibitem{Xing2011} Z.Z. Xing, Phys. Lett. B {\bf 696}, 232 (2011).

\bibitem{Tanimoto} M. Fukugita, M. Tanimoto, and T. Yanagida,
Phys. Rev. D {\bf 57}, 4429 (1998);
M. Tanimoto, Phys. Lett. B {\bf 483}, 417 (2000);
G.C. Branco and J.I. Silva-Marcos, Phys. Lett. B {\bf 526}, 104 (2002).

\bibitem{FX04} H. Fritzsch and Z.Z. Xing, Phys. Lett. B {\bf 598}, 237 (2004);
W. Rodejohann and Z.Z. Xing, Phys. Lett. B {\bf 601}, 176 (2004);
Z.Z. Xing, D. Yang, and S. Zhou, Phys. Lett. B {\bf 690}, 304 (2010).

\bibitem{FX99} H. Fritzsch and Z.Z. Xing, Prog. Part. Nucl. Phys. {\bf 45},
1 (2000); and references therein.

\bibitem{Xing03} Z.Z. Xing, J. Phys. G {\bf 29}, 2227 (2003).

\bibitem{Zhang} H. Zhang and Z.Z. Xing, Eur. Phys. J. C {\bf 41},
143 (2005); Z.Z. Xing and H. Zhang, Phys. Lett. B {\bf 618}, 131 (2005);
AIP Conf. Proc. {\bf 981}, 190 (2008).

\bibitem{Araki} See, e.g.,
T. Araki, C.Q. Geng, and Z.Z. Xing, arXiv:1012.2970;
accepted for publication in Phys. Lett. B.

\bibitem{RGE} See, e.g.,
S. Antusch, J. Kersten, M. Lindner,
and M. Ratz, Phys. Lett. B {\bf 544}, 1 (2002); S. Antusch and M. Ratz,
JHEP {\bf 0211}, 010 (2002);
J.W. Mei and Z.Z. Xing, Phys. Rev. D {\bf 70}, 053002 (2004);
Phys. Lett. B {\bf 623}, 227 (2005);
J.W. Mei, Phys. Rev. D {\bf 71}, 073012 (2005);
S. Luo and Z.Z. Xing, Phys. Lett.
B {\bf 632}, 341 (2006);
S. Goswami, S.T. Petcov, S. Ray,
W. Rodejohann, Phys. Rev. D {\bf 80}, 053013 (2009);
J. Bergstrom, T. Ohlsson, and H. Zhang, Phys. Lett. B {\bf 698}, 297 (2011).

\bibitem{Xing08} Z.Z. Xing, Phys. Rev. D {\bf 78}, 011301 (2008).

\end{thebibliography}
\end{document}